%% file: make_astro.tex
\documentclass{mpe_report}

\usepackage{psfig,graphicx,epsfig}
\usepackage{color}
\usepackage{amsmath,amssymb,epic,eepic,array}
\usepackage{natbib}

\begin{document}

\pagenumbering{arabic}
\setcounter{page}{181}

\renewcommand{\FirstPageOfPaper }{181}\renewcommand{\LastPageOfPaper }{184}\include{./mpe_report_schaffner}  \clearpage

\end{document}

%% file: mpe_report_schaffner.tex
\title{Neutron stars and quark stars: Two coexisting families of compact
  stars?} 

\author{J. Schaffner-Bielich}  
\institute{
Institut f\"ur Theoretische Physik/Astrophysik,
J. W. Goethe Universit\"at, Max-von-Laue Stra\ss e~1,
D--60438 Frankfurt am Main, Germany}

\maketitle

\begin{abstract}
  The mass-radius relation of compact stars is discussed with relation
  to the presence of quark matter in the core. The existence of a new
  family of compact stars with quark matter besides white dwarfs and
  ordinary neutron stars is outlined.
\end{abstract}


\section{Introduction}

Exotic compact stars dubbed quark stars have been a topic not only of
scientific but also of public interest, at least since the press release
of NASA of 2002 on the possible existence of compact stars with
unusually small radii and rapid cooling due to the presence of quark
matter in the dense interior.
\begin{figure}
\centerline{\psfig{file=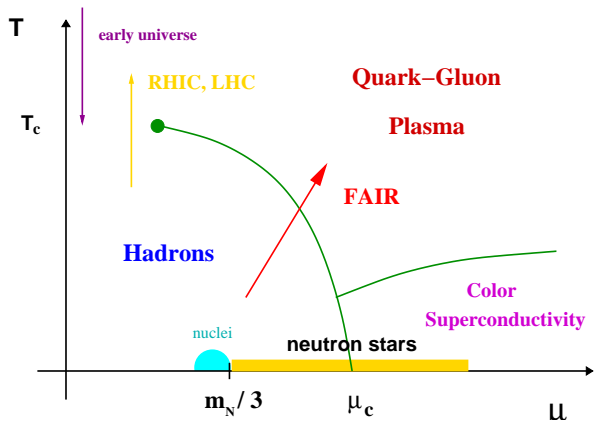,width=7.8cm,clip=}}
\caption{The phase diagram of strongly interacting matter (QCD).}
\label{fig:phasediagram}
\end{figure}
Indeed, the fundamental theory of strong interactions, quantum
chromodynamics (QCD), predicts that at large densities and/or
temperatures a phase transition to a plasma of (free) quarks and gluons
occurs. The phase diagram of QCD is depicted schematically in
Fig.~\ref{fig:phasediagram}. The early universe passed through that
phase diagram at (nearly) zero density and high temperatures. Neutron
star matter, on the other hand, is located at comparably small
temperatures and high densities. Interestingly, a first order phase
transition is likely to be present at high densities as present in the
core of neutron stars. This phase transition will be probed by the
collision of heavy ions at GSI, Darmstadt with the future international
facility FAIR.

\section{Quark matter and quark stars}

\begin{figure}
\centerline{\psfig{file=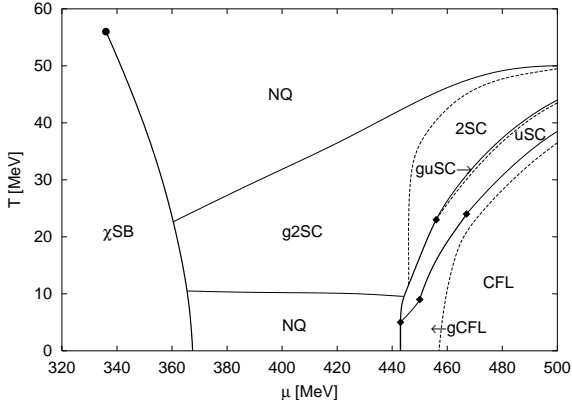,width=7.8cm,clip=}}
\caption{The possible phases in dense and hot quark matter (taken from
  \citet{Ruster:2005jc}).}.
\label{fig:csc_phases}
\end{figure}

Quark matter at high densities is remarkably rich in structure due to
the strong effects of pairing quarks to diquarks, the phenomenon of
colour-superconductivity.  The possible phases have been investigated
selfconsistently recently for cold neutron star matter, as well
as for hot, neutrino-rich proto-neutron stars, by
\citet{Ruster:2005jc,Ruster:2005ib} and are shown in
Fig.~\ref{fig:csc_phases}. Note, that the phase boundaries have been
determined by considering $\beta$-equilibrium.  The labels in the figure
stand for normal (unpaired) quark matter (NQ), two-flavour colour
superconducting phase (2SC), gapless 2SC phase (g2SC, etc.),
colour-flavour locked phase (CFL), gapless CFL phase (gCFL), which differ
in their quark pairing pattern. The phase transition relevant for
neutron stars is of first order due to symmetry arguments.  Matter
changes from the chirally broken phase ($\chi$SB), massive quarks, to
the (approximate) chirally restored phase with nearly massless quarks.
The corresponding order parameter is the (chiral) quark condensate,
which generates e.g.\ the mass of the nucleon besides the small current
quark masses.

As a model for cold and dense QCD matter, one can consider basically two
possibilities for a first-order chiral phase transition: A weakly
first-order chiral transition (or no true phase transition) or a
strongly first-order chiral transition. In the former case, there will
be just one type of compact star (ordinary neutron stars), in the
latter case {\em two} types of compact stars, as there exists a new stable
and more compact solution with smaller radii.

The essential ingredient is the relation between the pressure and the
energy density of the high-density (quark) matter. Usually, the MIT bag
model is adopted which assumes that free quarks are confined inside a
bag characterised by a vacuum energy density, the MIT bag constant.
Interactions up to order $\alpha_s^2$ of the strong coupling constant
have been reconsidered recently by \citet{FPS01} using perturbative QCD
and applied to quark stars. Interestingly, higher order corrections seem
to mimic the effect of the vacuum energy density of the MIT bag model so
that the mass-radius curves of both approaches are surprisingly similar.
For one particular choice of one free parameter, the maximum mass and
corresponding radius are $M_{\rm max} = 1.05\, M_\odot$ and $R_{\rm max}
= 5.8$ km with a central density of $n_{\rm max} = 15\,n_0$. For another
one, the values are $M_{\rm max} = 2.14\, M_\odot$, $R_{\rm max} = 12$
km, and $n_{\rm max} = 5.1\,n_0$. The latter case corresponds to
absolutely stable strange quark matter, i.e.\ to so called strange
(quark) stars \citep{Witten84,Haensel86,Alcock86}. Ordinary neutron stars
can collapse to this energetically favoured state. Note, that the
maximum mass of strange stars is comparable to ordinary stars, in fact it
is even slightly larger than two solar masses. In the former case,
quark matter is metastable and a surface layer of nuclear matter with a
mixed phase is present, which constitutes a so called hybrid (quark)
star.

There are a few other nonperturbative approaches on the market which
consider pure quark stars within an equation of state of dense QCD
matter using e.g.\ a Schwinger--Dyson model \citep{Blaschke99}, massive
quasi-particles \citep{Peshier00}, the Nambu--Jona-Lasinio model
\citep{Hanauske01}, hard-dense loop resummation schemes
\citep{Andersen:2002jz} and effects from colour superconductivity
\citep{Ruster04}. Historically, the first pure quark star was calculated by
\citet{Itoh70} even before the discovery of asymptotic freedom of QCD.

For hybrid stars, one has to match the quark phase to the low-density
hadronic phase. There are in principle two possible scenarios. The
pressure in the hadronic phase rises strongly with the baryochemical
potential (or baryon density) so that it
smoothly matches onto that of the quark phase and the phase transition
is weakly first order or a cross-over. Or the pressure increases slowly
with the baryochemical potential so that there is a large mismatch in
the slope of the hadronic and the quark phase and the phase transition is
strongly first order. Interestingly, there are hints that 
asymmetric matter up to $\sim 2 n_0$ has a relatively small pressure
which points to the latter scenario \citep{Akmal98}.  

\begin{figure}
\centerline{\psfig{file=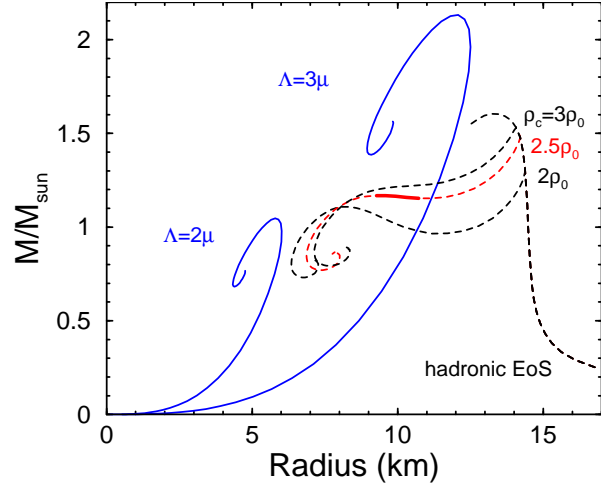,width=7.8cm,clip=} }
\caption{The mass-radius relation of pure quark stars in perturbative
  QCD (solid lines) and neutron stars and hybrid stars (dashed lines)
  taken from \citet{FPS01}.}
\label{fig:mr_quarkhadron}
\end{figure}

A strong first order phase transition in compact stars is of particular
interest as it allows for compact star twins. Various mass-radius
relation for the different scenarios outlined above are depicted in
Fig.~\ref{fig:mr_quarkhadron}. The solid lines show the mass-radius
relation for strange stars, which are selfbound and not bound by gravity
so that the curve starts at vanishing small masses.  The dashed lines
stand for the mass-radius relations of ordinary (hadronic) neutron stars
and hybrid stars, where the curve starts at some small mass at large
radii. These compact stars are bound by gravity. If a phase transition
to quark matter occurs at some critical density, the curve drops off
changing its slope. For some smaller values of the onset density of
quark matter, the slope changes again at small radii signalling a new
stable branch of the mass-radius curve.  So, if the phase transition is
weak, there exists at most ordinary neutron stars with a quark core
(hybrid stars). If however, the phase transition is strong, a new stable
solution to the Tolman-Oppenheimer-Volkoff equation for compact stars
becomes possible with maximum masses around $M \sim 1\, M_\odot$ and
corresponding radii of $R \sim 6$ km. In this case, the quark phase
dominates as $n \sim 15\, n_0$ at the center and there is only a small
hadronic mantle. Note, that a pure quark phase can be also present in an
ordinary hybrid star and is not restricted to the branch of the new
family of compact stars.

\begin{figure}
\centerline{\psfig{file=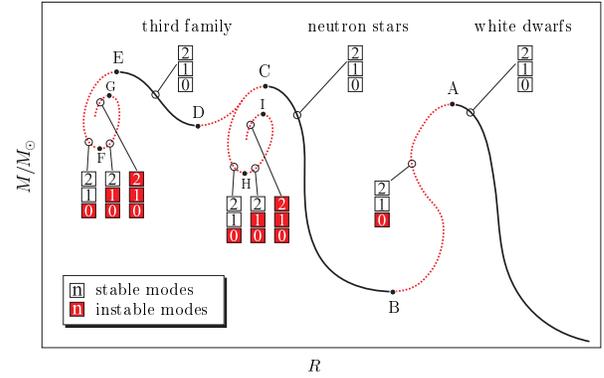,width=7.8cm,clip=} }
\caption{The schematic mass-radius relation of the three possible
  coexisting compact stars: white dwarfs, neutron stars and quark stars
  (taken from \citet{Schertler00}).}
\label{fig:twins}
\end{figure}

As there are two other families of compact stars known, white dwarfs and
neutron stars, the new solution constitutes a third family of compact
stars. The possibility of this new class of compact stars has been
outlined already by \citet{Gerlach68} and has been a subject of
discussion in the literature since then, see e.g.\ the historical notes
in \citet{JSBsqm04}. In the context of modern and thermodynamical
consistent description of the phase transition, the third family of
solutions was rediscovered by \citet{GK2000} and \citet{Schertler00} and
found its way into a standard textbook on compact stars
\citep{Glen_book}. The generic feature of the three families of compact
stars is outlined in Fig.~\ref{fig:twins}. The solid line from the right
to point A stands for the first family, white dwarfs, which are stabilised
by the electron degeneracy pressure. Then follows an unstable branch,
until the neutron degeneracy pressure is high enough to support stable
neutron stars from point B to C. Finally, the quark pressure allows for
stable quark stars from point D to E, which are even more compact than
ordinary neutron stars. Note, that the phase transition in neutron stars
must be strongly first order to reach the new stable branch, otherwise
the mass-radius curve terminates already in an unstable spiral along the
points H and I. It is also important to realize that {\em any} strong
first order phase transition will lead to a new stable branch of
solutions, as the only input to the Tolman-Oppenheimer-Volkoff equation
is the equation of state, irrespective of the underlying micro-physics
(see e.g.\ \citet{Macher05} for a purely parametric analysis).
Indeed, the new solution has been found in a large variety of different
approaches to the high-density equation of state, involving quark matter
in the MIT bag model \citep{GK2000,Mishustin03}, massive quasi-particles
of quarks \citep{Schertler00}, interacting quarks in perturbative QCD
\citep{FPS01}, Kaon condensation \citep{Banik01}, hyperon matter
\citep{Scha02}, and colour-superconducting quarks \citep{Banik03}.

\section{Signals for quark stars}

The existence of exotic matter in the core, a strong first order phase
transition and the third family of compact stars can be tested by
astronomical observations. We flash some of the many signals suggested in
the following.

Certainly, the mass-radius of pulsars and non-pulsating compact stars
allows for a unique test for the presence of a third family, which is
related to the so called phenomenon of the rising twins
\citep{Schertler00}. Standard massive neutron stars have a {\em smaller}
radius with increasing mass.  On the other hand, a new stable branch
allows for a pair of compact stars having the exact opposite behaviour,
i.e.\ $M_1<M_2$ with $R_1<R_2$!  A sizable mixed phase will lead to a
spontaneous spin-up of pulsars \citep{Glen97} as the pulsars passes
through the mixed phase during the standard spin-down evolution.  The
presence of exotic matter in general, be it strange quark matter, kaon
condensation or hyperon matter, causes a delayed collapse of a
proto-neutron star to a black hole \citep{Pons01L} as the hot
proto-neutron star allows for a higher maximum mass compared to the cold
(exotic) neutron star.  The collapse of a neutron star to the third
family can lead to a catastrophic rearrangement thereby emitting
gravitational waves, $\gamma$-rays, and a burst of neutrinos, see e.g.\
\citep{Mishustin03,Lin:2005zd}. The gravitational waves emitted from
colliding neutron stars can exhibit features which are sensitive to the
underlying equation of state including quark matter, be it for hybrid
star \citep{Oechslin04} or strange star collisions
\citep{Limousin:2004vc}.

So far several types of compact stars have been discussed. Let us
summarise the similarities and difference. Neutron stars with an
ordinary hadronic mantle and quark matter in the core are called hybrid
stars as there are two completely different phases in the interior of
the compact star. I used the term quark star for a special class of
hybrid stars which are located in a separate stable branch of the
mass-radius diagram constituting a third family of compact stars besides
white dwarfs and neutron stars. Those quark stars have for sure a strong
first order phase transition and a pure exotic phase in the core, which
is most likely but not necessarily restricted to quark matter. Finally,
strange (quark) stars are selfbound stars consisting of absolutely
stable strange quark matter only. They are purely hypothetical as they
rely on the Bodmer-Witten hypothesis \citep{Bodmer71,Witten84} that
there is some exotic matter, strange quark matter, which is more bound
than nuclei, so that nuclei are transformed to strange quark matter with
releasing energy.

The hypothetical selfbound stars are characterised in general by a
vanishing pressure at a finite energy density and a mass-radius relation
starting at the origin (ignoring a possible nuclear crust) allowing for
arbitrarily small masses and radii down to nuclear scales. 

On the contrary, ordinary neutron stars and hybrid stars are bound by
gravity where matter has a finite pressure for all energy densities.
The corresponding mass-radius relation starts at large radii with a
minimum neutron star mass of $M\sim 0.1M_\odot$ at $R\sim 200$ km. This
minimum neutron star mass is dictated by the properties of the
degenerate low-density gas of neutrons, electrons and nuclei. Note, that
white dwarfs can have much smaller masses due to the formation of a
lattice of nuclei surrounded by free electrons. The lattice energy for
neutron stars is only marginally, if at all, able to support the compact
star \citep{BPS}.

Selfbound strange stars have similar maximum masses and radii, with a
similar nuclear crust, etc. But there are some features which are unique
for strange stars. For example, they can have extremely small masses
with small radii. White dwarfs with a core of strange star material can
exist with unusual mass-radius relation
\citep{Glendenning:1994zb,GKW95}. And bare, hot strange stars allow for
super-Eddington luminosity as the material is bound by interactions and
not by gravity (see e.g., \citet{Page:2002bj}).

\section{Summary}

Our present understanding of dense matter in strong interactions opens
the possibility of having a first order phase transition from ordinary
matter to some exotic (quark) matter at high densities which is
connected to chiral symmetry restoration.  If the phase transition is
strongly first order it generates a new, stable solution for compact
stars besides white dwarfs and neutron stars. Note, that the new stable
branch in the mass-radius diagram is not constraint by present
mass-radius data!  The transition to quark matter has not only impacts
on the global properties of neutron stars and pulsars, but also on
supernovae and proto-neutron star evolution and neutron star mergers.
The third family of compact stars, quark stars, coexist with ordinary
neutron stars, so it might be worthwhile to check for the possibility of
having two different classes of pulsar families!


\begin{acknowledgements}

I thank the organisers for their kind invitation and especially
Slavko Bogdanov,
Okkie de Jager,
Hovik Gregorian, 
Frank Haberl,
Matthias Hempel, 
Michael Kramer, 
Wolfgang Kundt, 
Harald Lesch, 
Aristeidis Noutsos,
Dany Page,
Mal Ruderman, 
Irina Sagert,
Morten Stejner, 
Joachim Tr\"umper, 
Roberto Turolla,
Christo Venter,  
Fridolin Weber, 
Slava Zavlin
for the numerous and enjoyable discussions during the meeting. 

\end{acknowledgements}
   
\bibliographystyle{aa}
\bibliography{all,literat}

%% file: make_astro.bbl
\begin{thebibliography}{33}
\expandafter\ifx\csname natexlab\endcsname\relax\def\natexlab#1{#1}\fi

\bibitem[{Akmal {et~al.}(1998)Akmal, Pandharipande, \& Ravenhall}]{Akmal98}
Akmal, A., Pandharipande, V.~R., \& Ravenhall, D.~G. 1998, Phys. Rev. C, 58,
  1804

\bibitem[{Alcock {et~al.}(1986)Alcock, Farhi, \& Olinto}]{Alcock86}
Alcock, C., Farhi, E., \& Olinto, A. 1986, Astrophys. J., 310, 261

\bibitem[{Andersen \& Strickland(2002)}]{Andersen:2002jz}
Andersen, J.~O. \& Strickland, M. 2002, Phys. Rev., D66, 105001

\bibitem[{Banik \& Bandyopadhyay(2001)}]{Banik01}
Banik, S. \& Bandyopadhyay, D. 2001, Phys. Rev. C, 64, 055805

\bibitem[{Banik \& Bandyopadhyay(2003)}]{Banik03}
Banik, S. \& Bandyopadhyay, D. 2003, Phys. Rev. D, 67, 123003

\bibitem[{Baym {et~al.}(1971)Baym, Pethick, \& Sutherland}]{BPS}
Baym, G., Pethick, C., \& Sutherland, P. 1971, Astrophys. J., 170, 299

\bibitem[{Blaschke {et~al.}(1999)Blaschke, Grigorian, Poghosyan, Roberts, \&
  Schmidt}]{Blaschke99}
Blaschke, D., Grigorian, H., Poghosyan, G., Roberts, C.~D., \& Schmidt, S.~M.
  1999, Phys. Lett. B, 450, 207

\bibitem[{Bodmer(1971)}]{Bodmer71}
Bodmer, A.~R. 1971, Phys. Rev. D, 4, 1601

\bibitem[{Fraga {et~al.}(2001)Fraga, Pisarski, \& Schaffner-Bielich}]{FPS01}
Fraga, E.~S., Pisarski, R.~D., \& Schaffner-Bielich, J. 2001, Phys. Rev. D, 63,
  121702(R)

\bibitem[{Gerlach(1968)}]{Gerlach68}
Gerlach, U.~H. 1968, Phys. Rev., 172, 1325

\bibitem[{Glendenning(2000)}]{Glen_book}
Glendenning, N.~K. 2000, Compact Stars --- Nuclear Physics, Particle Physics,
  and General Relativity, 2nd edn. (New York: Springer)

\bibitem[{Glendenning \& Kettner(2000)}]{GK2000}
Glendenning, N.~K. \& Kettner, C. 2000, Astron. Astrophys., 353, L9

\bibitem[{Glendenning {et~al.}(1995{\natexlab{a}})Glendenning, Kettner, \&
  Weber}]{Glendenning:1994zb}
Glendenning, N.~K., Kettner, C., \& Weber, F. 1995{\natexlab{a}}, Astrophys.
  J., 450, 253

\bibitem[{Glendenning {et~al.}(1995{\natexlab{b}})Glendenning, Kettner, \&
  Weber}]{GKW95}
Glendenning, N.~K., Kettner, C., \& Weber, F. 1995{\natexlab{b}}, Phys. Rev.
  Lett., 74, 3519

\bibitem[{Glendenning {et~al.}(1997)Glendenning, Pei, \& Weber}]{Glen97}
Glendenning, N.~K., Pei, S., \& Weber, F. 1997, Phys. Rev. Lett., 79, 1603

\bibitem[{Haensel {et~al.}(1986)Haensel, Zdunik, \& Schaeffer}]{Haensel86}
Haensel, P., Zdunik, J.~L., \& Schaeffer, R. 1986, Astron. Astrophys., 160, 121

\bibitem[{Hanauske {et~al.}(2001)Hanauske, Satarov, Mishustin, St{\"o}cker, \&
  Greiner}]{Hanauske01}
Hanauske, M., Satarov, L.~M., Mishustin, I.~N., St{\"o}cker, H., \& Greiner, W.
  2001, Phys. Rev. D, 64, 043005

\bibitem[{Itoh(1970)}]{Itoh70}
Itoh, N. 1970, Prog. Theor. Phys., 44, 291

\bibitem[{Limousin {et~al.}(2005)Limousin, Gondek-Rosinska, \&
  Gourgoulhon}]{Limousin:2004vc}
Limousin, F., Gondek-Rosinska, D., \& Gourgoulhon, E. 2005, Phys. Rev., D71,
  064012

\bibitem[{Lin {et~al.}(2006)Lin, Cheng, Chu, \& Suen}]{Lin:2005zd}
Lin, L.-M., Cheng, K.~S., Chu, M.~C., \& Suen, W.-M. 2006, Astrophys. J., 639,
  382

\bibitem[{Macher \& Schaffner-Bielich(2005)}]{Macher05}
Macher, J. \& Schaffner-Bielich, J. 2005, Eur. J. Phys., 26, 341

\bibitem[{Mishustin {et~al.}(2003)Mishustin, Hanauske, Bhattacharyya, Satarov,
  St{\"o}cker, \& Greiner}]{Mishustin03}
Mishustin, I.~N., Hanauske, M., Bhattacharyya, A., {et~al.} 2003, Phys. Lett.
  B, 552, 1

\bibitem[{Oechslin {et~al.}(2004)Oechslin, Ury{\=u}, Poghosyan, \&
  Thielemann}]{Oechslin04}
Oechslin, R., Ury{\=u}, K., Poghosyan, G., \& Thielemann, F.~K. 2004, Mon. Not.
  R. Astron. Soc., 349, 1469

\bibitem[{Page \& Usov(2002)}]{Page:2002bj}
Page, D. \& Usov, V.~V. 2002, Phys. Rev. Lett., 89, 131101

\bibitem[{Peshier {et~al.}(2000)Peshier, K\"ampfer, \& Soff}]{Peshier00}
Peshier, A., K\"ampfer, B., \& Soff, G. 2000, Phys. Rev. C, 61, 045203

\bibitem[{Pons {et~al.}(2001)Pons, Steiner, Prakash, \& Lattimer}]{Pons01L}
Pons, J.~A., Steiner, A.~W., Prakash, M., \& Lattimer, J.~M. 2001, Phys. Rev.
  Lett., 86, 5223

\bibitem[{R{\"u}ster \& Rischke(2004)}]{Ruster04}
R{\"u}ster, S.~B. \& Rischke, D.~H. 2004, Phys. Rev. D, 69, 045011

\bibitem[{R{\"u}ster {et~al.}(2005)R{\"u}ster, Werth, Buballa, Shovkovy, \&
  Rischke}]{Ruster:2005jc}
R{\"u}ster, S.~B., Werth, V., Buballa, M., Shovkovy, I.~A., \& Rischke, D.~H.
  2005, Phys. Rev., D72, 034004

\bibitem[{R{\"u}ster {et~al.}(2006)R{\"u}ster, Werth, Buballa, Shovkovy, \&
  Rischke}]{Ruster:2005ib}
R{\"u}ster, S.~B., Werth, V., Buballa, M., Shovkovy, I.~A., \& Rischke, D.~H.
  2006, Phys. Rev., D73, 034025

\bibitem[{Schaffner-Bielich(2005)}]{JSBsqm04}
Schaffner-Bielich, J. 2005, J. Phys. G, 31, S651

\bibitem[{Schaffner-Bielich {et~al.}(2002)Schaffner-Bielich, Hanauske,
  St{\"o}cker, \& Greiner}]{Scha02}
Schaffner-Bielich, J., Hanauske, M., St{\"o}cker, H., \& Greiner, W. 2002,
  Phys. Rev. Lett., 89, 171101

\bibitem[{Schertler {et~al.}(2000)Schertler, Greiner, Schaffner-Bielich, \&
  Thoma}]{Schertler00}
Schertler, K., Greiner, C., Schaffner-Bielich, J., \& Thoma, M.~H. 2000, Nucl.
  Phys., A677, 463

\bibitem[{Witten(1984)}]{Witten84}
Witten, E. 1984, Phys. Rev. D, 30, 272

\end{thebibliography}
